\documentclass[prb,twocolumn,superscriptaddress]{revtex4}

\usepackage{amsmath}
\usepackage{amssymb}
\usepackage{bm}
\usepackage{epsfig}
\usepackage{graphicx}
\usepackage{color}

\begin{document}

\title{Kinetic Monte Carlo Approach to Non-equilibrium Bosonic Systems}

\author{T. C. H. Liew}
\affiliation{Division of Physics and Applied Physics, School of Physical and Mathematical Sciences, Nanyang Technological University, 21 Nanyang Link, Singapore 637371}

\author{H. Flayac}
\affiliation{Institute of Theoretical Physics, Ecole Polytechnique F\'ed\'erale de Lausanne (EPFL), CH-1015 Lausanne, Switzerland}

\author{D. Poletti}
\affiliation{Singapore University of Technology and Design, 8 Somapah Road, 487372 Singaore}

\author{I. G. Savenko}
\affiliation{Center for Theoretical Physics of Complex Systems, Institute for Basic Science (IBS), Daejeon 34051, Republic of Korea}
\affiliation{Nonlinear Physics Centre, Research School of Physics and Engineering, The Australian National University, Canberra ACT 2601, Australia}
\affiliation{ITMO University, St. Petersburg 197101, Russia}

\author{F. P. Laussy}
\address{Russian Quantum Center, Novaya 100, 143025 Skolkovo, Moscow Region, Russia}
\address{Departamento de F\'isica Te\'orica de la Materia Condensada and Condensed Matter Physics Center (IFIMAC), Universidad Aut\'onoma de Madrid, E-28049, Spain}

\begin{abstract}
We consider the use of a Kinetic Monte Carlo approach for the description of non-equilibrium bosonic systems, taking non-resonantly excited exciton-polariton condensates and bosonic cascade lasers as examples. In the former case, the considered approach allows the study of the cross-over between incoherent and coherent regimes, which represents the formation of a quasi-condensate that forms purely from the action of energy relaxation processes rather than interactions between the condensing particles themselves. In the latter case, we show that a bosonic cascade can theoretically develop an output coherent state.
\end{abstract}

\date{\today}

\maketitle

\section{Introduction}

Although Bose-Einstein condensation (BEC) was originally defined as an effect taking place in thermal equilibrium, it is striking to see that the concept has been generalized to nonequilibrium systems. For example, the BEC of photons in a cavity~\cite{Klaers2010} has been reported and several groups have studied the BEC of exciton-polaritons (hybrid light-matter quasiparticles) appearing in semiconductor microcavities~\cite{Kasprzak2006,Balili2007,Lai2007}. Here BEC is characterized~\cite{Byrnes2014} by the spontaneous formation of coherence, typically measured by the transition of the second order coherence function with increasing particle density, as reported by several experimental groups~\cite{Love2008,Kasprzak2008,Horikiri2010,Tempel2012,RahimiIman2012,Amthor2015,Kim2016}.

The physics of nonequilibrium condensates has been shown to be radically different from that of equilibrium systems, where condensates may form in non-ground states~\cite{Maragkou2010}, multiple states~\cite{Vorberg2013}, and have distributions undescribable by a single temperature.

The theoretical description of nonequilibrium condensates typically requires an explicit treatment of energy relaxation processes. Such processes compete with dissipative processes, which cause particles to be lost from the system before reaching the ground state. The dynamical interplay of relaxation and dissipation ultimately determines the steady-state of the system. Energy relaxation mechanisms have been handled previously in exciton-polariton systems using semiclassical Boltzmann equations~\cite{Porras2002,Doan2005,Kasprzak2008b,Cao2008} or introduced phenomenologically into mean-field equations~\cite{Read2009,Wouters2010,Wouters2012,Sieberer2013}. Methods treating energy relaxation from first principles have also been developed based on stochastic sampling of mean-field equations~\cite{Savenko2013} or their hybridization with the Boltzmann equations~\cite{Solnyshkov2014}. However, these methods do not account for quantum fluctuations, which are needed for the unified treatment of nonequilibrium condensation below and above threshold. In principle, density matrix approaches~\cite{Racine2014} (possibly supplemented with Monte Carlo techniques~\cite{Molmer1993}) are applicable to this task, however, in practice they are only feasible for systems with small numbers of particles and modes~\cite{Flayac2015}. Bosonic cascade lasers~\cite{Liew2013,Tzimis2015}, which may operate with millions of particles have been treated with stochastic sampling of the positive-P distribution~\cite{Chaturvedi1977,Liew2016}, however, such a method is only accurate in the presence of an initial coherent state.

In the present work we employ a kinetic Monte Carlo approach based on quantum Boltzmann equations for the description of non-equilibrium multimode open quantum systems. Kinetic Monte Carlo has been developed under different names in different fields, from vacancy migration in binary ordered alloys~\cite{Young1966}, the Ising model~\cite{Bortz1975} and chemical reactions~\cite{Gillespie1976}. A good overview of the method can be found in Ref.~\onlinecite{Fichthorn1991}. The approach allows stochastic sampling of the quantum particle distribution function and allows the treatment of systems with up to hundreds of modes with possibly thousands of particles each. From the particle distribution functions we have full access to the coherence statistics, as characterized by the second order correlation function. We apply the technique to two specific examples: polariton condensation in one-dimensional microwires and terahertz lasing in bosonic cascade lasers.

In the former case we are able to describe the gradual cross-over from incoherent population of excited states to partial coherence in non-ground states and the formation of a fully coherent BEC with increasing particle density. The responsible energy relaxation processes are described from first principles, accounting for polariton-phonon scattering and the scattering of polaritons with hot exciton states~\cite{Porras2002}. Aside these interaction processes, it is notable that additional interactions between the condensing particles themselves are not required for the formation of a condensate (which is consistent with the original equilibrium theory of BEC of the ideal gas).

In the case of bosonic cascade lasers, we access for the first time theoretically the coherence of the lasing mode and show that it can be useful for terahertz lasing with high quantum efficiency.

\section{Generic Kinetic Monte Carlo Approach}

We start with the consideration of a set of $M$ discrete modes with populations $n_1$, $n_2$, $\ldots$, $n_M$. The probability of the system being in any particular state at time $t$ is $P_{n_1,n_2,\ldots,n_M}(t)$. The probability distribution contains sufficient information to calculate the quantum expectation values of a variety of quantities, in particular those with operators that commute with the number operator. For example, one can calculate:
\begin{align}
\langle n_i(t) \rangle &= \sum P_{n_1,n_2,\ldots,n_i,\ldots,n_M}(t) n_i\\
\langle n_i^2(t) \rangle &= \sum P_{n_1,n_2,\ldots,n_i,\ldots,n_M}(t) n_i^2
\end{align}
Of course the probability distribution $P_{n_1,n_2,\ldots,n_M}(t)$ does not contain all the information on the state of the system, which would require the full quantum density matrix. However, from the above we can gain access to the second order correlation function $g_{2,n_i}(t)=\langle n_i^2(t) \rangle/\langle n_i(t) \rangle^2$, which is the parameter typically used to measure the coherence of a given mode. The above prescription can also be easily generalized to the case of non-zero time delay and cross-correlations between different modes.

The calculation of the quantum probability distribution can be based on the quantum Boltzmann master equation, with generic form:

\begin{align}
&\frac{dP_{n_1,n_2,\ldots,n_M}(t)}{dt}\notag\\
&\hspace{5mm}=\sum_{ij} W_{i\rightarrow j}P_{n_1,n_2,\ldots,n_j-1,n_i+1,\ldots,n_M}(t)\left(n_i+1\right)n_j\notag\\
&\hspace{5mm}+\sum_i\frac{1}{\tau_i}\left[P_{n_1,n_2,\ldots,n_i+1,\ldots,n_M}(t)\left(n_i+1\right)\right.\notag\\
&\hspace{30mm}\left.-P_{n_1,n_2,\ldots,n_i,\ldots,n_M}(t)n_i\right]\notag\\
&\hspace{5mm}+\sum_i\Gamma_i\left[P_{n_1,n_2,\ldots,n_i-1,\ldots,n_M}(t)n_i\right.\notag\\
&\hspace{30mm}\left.-P_{n_1,n_2,\ldots,n_i,\ldots,n_M}(t)\left(n_i+1\right)\right]\label{eq:QuantumBoltzmann}
\end{align}

The first term represents stimulated scattering processes between modes, where $W_{i\rightarrow j}$ is the bare (spontaneous) scattering rate from mode $i$ to mode $j$. In an exciton-polariton system, this term would include phonon emission (or absorption) processes as well as scattering processes involving hot excitons~\cite{Porras2002}. These processes introduce a temperature dependence of the system via the temperatures of phonon or exciton baths.

The second and third terms represent decay and incoherent/non-resonant pumping of the modes, at rates $\tau_i$ and $\Gamma_i$, respectively. Their form is consistent with the Liouvillian operator for the full quantum density matrix, written in Ref.~\onlinecite{Laussy2008} for the case of incoherent/non-resonant pumping. In principle other scattering processes (e.g., parametric scattering processes~\cite{Savvidis2000}) can also be included, where the generic form is the coupling of one probability in the distribution to another.

Equation~\ref{eq:QuantumBoltzmann} can be simulated numerically given $W_{i\rightarrow j}$, $\tau_i$, and $\Gamma_i$ using kinetic Monte Carlo. This approach is based on first defining an initial state:
\begin{equation}
\left(n_1,n_2,\ldots,n_M\right)
\end{equation}
Equation~\ref{eq:QuantumBoltzmann} defines the scattering rates to other possible states to which the above state can jump to. A probability distribution of possible jumps to other states is associated to the scattering rates and a random quantum jump is selected from the probability distribution. The jump time defines the amount of time the system spends in the original state, from which the calculation of expectation values can be updated. Then the process is repeated until the end of the time range of the calculation. The process is then further repeated sampling over different quantum trajectories characterized by different stochastic quantum jumps. The system is able to attain a steady state, characterized by constant average expectation values. This is because nonlinear loss processes~\cite{Keeling2008} have effectively been accounted for in the quantum Boltzmann equations; when a given mode becomes highly occupied, the probability for it to lose particles increases such that its occupation is bounded.

Since we neglect off-diagonal elements in the density matrix, we note that our approach is strictly speaking valid only when the system is not too far above the condensation threshold and the various scattering processes can be obtained accurately from single-particle wavefunctions. Far above threshold, polaritons are typically modelled with the mean-field Gross-Pitaevskii equation, where coherence is assumed. Here, we are interested in the behavior crossing the threshold. In principle, polariton-polariton scattering could renormalize the energy dispersion and alter the various energy relaxation rates, however, this does not affect the general trend of relaxing to the lowest available state, which is why we will obtain results consistent with experiments even above threshold.

Given that the range of validity of the quantum Boltzmann approach is the same as that for standard classical Boltzmann equations, the quantum Boltzmann equations can always be reduced to classical ones, depending on the quantities of interest. Our motivation for working with quantum Boltzmann equations is that they give access to second order correlations.

\section{Non-Equilibrium Condensation in Polariton Microwires}

Exciton-polariton systems are short-lived bosonic quantum systems that are subject to weak energy relaxation processes. Consequently they are an exemplary non-equilibrium quantum system. They have been experimentally shown to form Bose-Einstein condensates~\cite{Kasprzak2006,Balili2007,Lai2007}, yet they may also become trapped in non-ground states~\cite{Richard2005} or form non-ground state condensates~\cite{Maragkou2010}.

\begin{figure}[h!]
\includegraphics[width=\columnwidth]{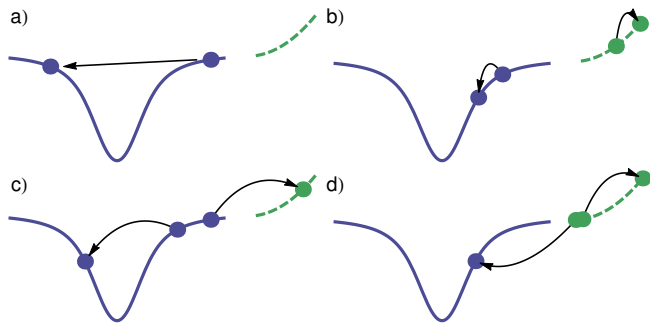}
\caption{(color online) Polariton energy relaxation mechanisms illustrated on the low-momentum polariton dispersion (blue/solid). The green/dashed curve illustrates the high-momentum exciton dispersion. a) Polariton-phonon scattering. b) Polariton-exciton scattering. c) Polariton-polariton to polariton-exciton scattering. d) Polariton pumping.}
\label{fig:ScatteringMechanisms}
\end{figure}

For simplicity, we will consider a one-dimensional exciton-polariton system or microwire~\cite{Wertz2010}. The study of partial energy relaxation processes in such systems is particularly relevant to the study of polariton condensate transistors~\cite{Anton2012,Anton2013} and the control of spin currents for spintronics~\cite{Anton2015,Gao2015}. The main energy relaxation mechanisms in this system arise from polariton scattering with acoustic phonons~\cite{Piermarocchi1996} and the scattering of polaritons with high momentum exciton states that can be considered as a reservoir~\cite{Porras2002} (provided that we are not interested in the coherence statistics of these excitons). The scattering processes are illustrated in Fig.~\ref{fig:ScatteringMechanisms}.

\begin{figure}[h!]
\includegraphics[width=\columnwidth]{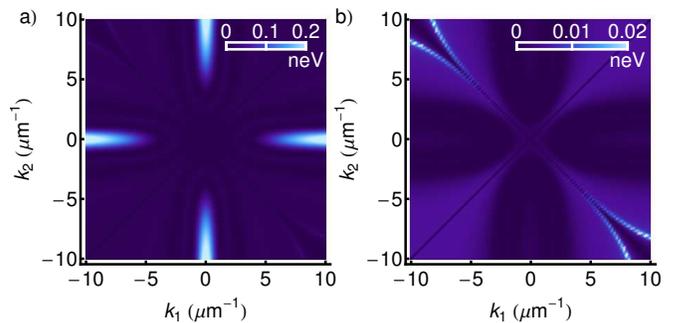}
\caption{(color online) Polariton-phonon scattering rates from mode $k_1$ to $k_2$. a) Polariton relaxation processes. The scattering to $k=0$ modes is larger than that of other modes due to the larger density of states at $k=0$ in a 1D system. b) Polariton excitation processes, requiring non-zero temperature.}
\label{fig:PhononScattering}
\end{figure}

The calculation of the polariton-phonon scattering rates is shown in the Appendix. For typical parameters we find the result shown in Fig.~\ref{fig:PhononScattering} for a temperature of $5$K. Due to the reduced density of states in a one-dimensional system, as compared to planar two-dimensional microcavities, we find that the polariton-phonon scattering rates are small, below neV. Given that typical polariton decay rates are at least $10\mu$eV in typical microcavities, {\it polariton-phonon scattering alone is insufficient to describe the relaxation of polaritons in one-dimensional systems}. Even accounting for bosonic stimulation, very large polariton occupation numbers or very high lifetime microcavities~\cite{Sun2017} would be needed to make polariton-phonon scattering dominant.

For this reason, it is important to account for the polariton-exciton scattering processes illustrated in Figs.~\ref{fig:ScatteringMechanisms}b-d. The calculation of these rates is outlined in the Appendix. The process in Fig.~\ref{fig:ScatteringMechanisms}b adds to the scattering rates $W_{ij}$ introduced in Eq.~\ref{eq:QuantumBoltzmann}. The processes in Fig.~\ref{fig:ScatteringMechanisms}c require the addition of new terms in the quantum Boltzmann equation (Eq.~\ref{eq:QuantumBoltzmann}) of the form:
\begin{equation}
W_{ij\rightarrow l}P_{n_1,\ldots,n_l-1,\ldots,n_i+1,\ldots,n_j+1,\ldots,n_M}(t)(n_i+1)(n_j+1)n_l
\end{equation}
We assume that the incoherent pumping processes of the system can be derived primarily from the process in Fig.~\ref{fig:ScatteringMechanisms}d to provide $\Gamma_i$. It should be noted that the rates of the pumping processes and the scattering process illustrated in Fig.~\ref{fig:ScatteringMechanisms}b are proportional to the occupation of exciton reservoir states, which is modelled with a Boltzmann distribution of the form:
\begin{equation}
n_{ex,k}=n_{ex}e^{-\hbar^2k^2/(2m_Xk_BT)},
\end{equation}
where $m_X$ is the exciton effective mass (taken as 0.22 times the free electron mass in GaAs based microcavities) and $n_{ex}$ is a parameter representing the maximum occupation of a state in the thermal exciton reservoir. This parameter can be taken as a measure of the strength of incoherent pumping in the system, which would be controlled experimentally via the intensity of a non-resonant laser or current from an electrical injection mechanism.

\begin{figure}[h!]
\includegraphics[width=\columnwidth]{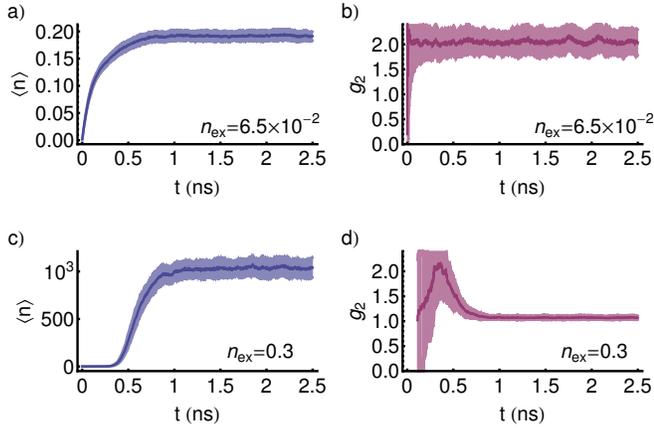}
\caption{(color online) a) Time dependence of $\langle n(t)\rangle$ of the ground state below threshold. b) Corresponding time dependence of $g_2(t)$. c) Time dependence of $\langle n(t)\rangle$ for the ground state ($k=0$) above threshold. d) Corresponding time dependence of $g_2(t)$. In each panel the light shaded region represents the statistical error, corresponding to one standard deviation.}
\label{fig:MicrowireTimeDependence}
\end{figure}

Given the aforementioned scattering rates for typical microcavity parameters (given in the Appendix) we obtain the time dependence of the average occupation of the ground state below threshold shown in Fig.~\ref{fig:MicrowireTimeDependence}a. Here the system is evolved from an initial vacuum state. The average occupation remains below unity and the state is that of an incoherent state with $g_2=2$ (Fig.~\ref{fig:MicrowireTimeDependence}b). The shown quantities are here obtained after averaging over more than $10^9$ quantum jumps. By studying the statistical variation over different runs, one can obtain an estimate for the error in the obtained quantities. Since for larger occupations one has to sample a larger number of different states, the statistical error in the occupations is larger.

Above threshold, we find a large occupation of the ground state developing after an initial stabilization time, as shown in Fig.~\ref{fig:MicrowireTimeDependence}c. This is accompanied by the formation of coherence characterized by $g_2=1$, as shown in Fig.~\ref{fig:MicrowireTimeDependence}d. While the statistical error in $g_2$ is very large when the occupation numbers are small, which correspond to a small denominator in calculating $g_2$ and consequently large effects of small fluctuations in $\langle n\rangle$, the statistical error above threshold becomes small and indistinguishable in the plot.

We stress that while we have accounted for polariton-phonon scattering, it does not affect significantly our results, in which polariton-exciton scattering is the dominant and sufficient energy relaxation mechanism. This is consistent with earlier works~\cite{Porras2002,Flayac2015}.

\begin{figure}[h!]
\includegraphics[width=\columnwidth]{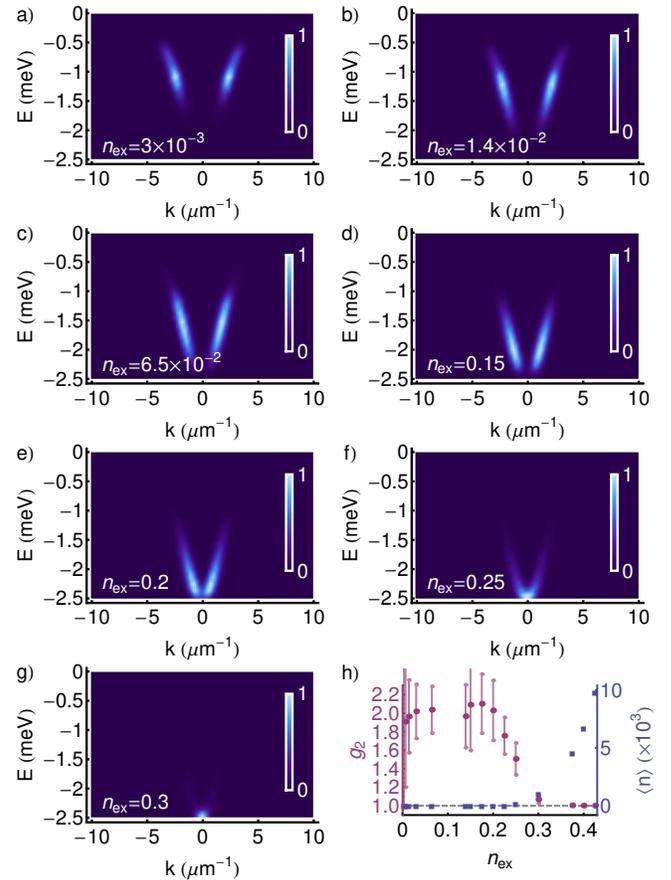}
\caption{(color online) a-g) Momentum distribution of polaritons in the long-time (continuous wave excitation) limit for increasing pumping intensity, represented by the maximum reservoir occupation number, $n_{ex}$. h) Dependence of the second order correlation function, $g_2$ (circles and left-hand scale), and the average occupation number, $\langle n\rangle$ (squares and right-hand scale), for the highest occupied state as a function of the reservoir occupation number. While there is a significant statistical error below threshold, the error bars narrow upon condensation.}
\label{fig:MicrowirePowerDependence}
\end{figure}

In addition to describing the behaviour of polaritons below and above threshold, the kinetic Monte Carlo theory is able to access the cross-over between the incoherent and condensed regimes. Figure~\ref{fig:MicrowirePowerDependence} illustrates the change in the momentum distribution of polaritons on the polariton dispersion. As the pumping intensity is increased the various energy relaxing scattering mechanisms become more and more stimulated. This is captured as a gradual overcoming of the bottleneck region~\cite{Richard2005}, before full condensation is obtained at large pumping strength. Here the large majority of polaritons collect in the system ground state and full coherence is characterized by $g_2=1$.

It is worth noting that the obtained phenomenon of condensation is obtained here {\it without direct interactions between condensing particles}. The interactions that we introduce are only to provide a physical mechanism of energy relaxation, but in principle any mechanism of energy relaxation would generate similar behaviour. The considered system is thus a nonequilibrium analogue of the non-interacting ideal gas. It should be noted though that here we are considering a confined system, in which the quantized modes in $k$ correspond to the levels discretely modelled in our approach.

While we have focused on the second order coherence, polariton condensates are typically also characterized by the appearance of a first order spatial coherence, which decays exponentially with distance~\cite{Mouchliadis2008,Belykh2013}. As our technique neglects off-diagonal terms in the density matrix we are unable to access this property, which could be treated with other techniques~\cite{Sarchi2007,Doan2008}.

\section{Coherence Formation in Bosonic Cascade Lasers}

A bosonic cascade laser is composed of a series of equidistant energy levels and was originally proposed for the high efficiency generation of terahertz (THz) frequency radiation~\cite{Liew2013}. Potential realizations making use of parabolic quantum wells are in experimental development~\cite{Tzimis2015}. When a particle is excited in a particular level of the cascade it is assumed that it can undergo a radiative transition to the next level in the cascade. Thus, in the case that the radiative transition is at THz frequency, one can have a high quantum efficiency process where an optical quantum of energy injected into the system undergoes multiple energy relaxing processes resulting in the emission of many THz frequency photons. The THz emission processes are typically weak in strength, but they can become enhanced by bosonic final state stimulation at high occupation numbers. When the system is placed inside a THz cavity, it has been assumed that the result will be the generation of a coherent THz mode although the theory of such a process has not been attempted.

The bosonic cascade laser can be described by the quantum Boltzmann rate equations:
\begin{align}
&\frac{dP_{n_1,\ldots,n_M,n_T}}{dt}=\notag\\
&\hspace{5mm}P_0\sum_{\lambda=1}^M\left[P_{n_1,\ldots,n_\lambda-1,\ldots,n_M,n_T}-P_{n_1,\ldots,n_\lambda,\ldots,n_M,n_T}\right]\notag\\
&\hspace{2mm}+W\sum_{\lambda=2}^M\left[-P_{n_1,\ldots,n_{\lambda-1},n_\lambda,\ldots,n_M,n_T}n_{\lambda-1}(n_\lambda+1)n_T\right.\notag\\
&\hspace{5mm}+P_{n_1,\ldots,n_{\lambda-1}-1,n_\lambda+1,\ldots,n_M,n_T-1}n_{\lambda-1}(n_\lambda+1)n_T\notag\\
&\hspace{5mm}+P_{n_1,\ldots,n_{\lambda-1}+1,n_\lambda-1,\ldots,n_M,n_T+1}(n_{\lambda-1}+1)n_\lambda(n_T+1)\notag\\
&\hspace{5mm}\left.-P_{n_1,\ldots,n_{\lambda-1},n_\lambda,\ldots,n_M,n_T}(n_{\lambda-1}+1)n_\lambda(n_T+1)\right]\notag\\
&\hspace{2mm}+\frac{1}{\tau}\sum_{\lambda=1}^M\left[P_{n_1,\ldots,n_\lambda+1,\ldots,n_M,n_T}(n_\lambda+1)\right.\notag\\
&\hspace{10mm}-\left.P_{n_1,\ldots,n_\lambda,\ldots,n_M,n_T}n_\lambda\right]\notag\\
&\hspace{2mm}+\frac{1}{\tau_T}\left[P_{n_1,\ldots,n_M,n_T+1}(n_T+1)-n_TP_{n_1,\ldots,n_M,n_T}\right]
\end{align}
where $M$ bosonic levels have populations $n_1$, $n_2$, $\ldots$, $n_M$, and $n_T$ is the number of THz photons in the THz cavity. We assume for simplicity an equal pumping rate of all levels in the cascade by a coherent driving of strength $P_0$. $\tau$ is the decay rate of bosons in each level, $\tau_T$ the decay rate of THz photons, and $W$ is the nearest neighbour level scattering rate, which is assumed here independent of the level index for simplicity. Rather than repeating the detailed calculation of the scattering rate, we take the value of $W\tau=8.3\times10^{-7}$ consistent with Refs.~\onlinecite{Liew2013,Liew2016}. Since our main objective is to derive the formation of coherence, that is, lasing in the THz mode we will consider a fixed value of $\tau_T=0.1\tau$ rather than presenting a detailed dependence on all parameters.

\begin{figure}[h!]
\includegraphics[width=\columnwidth]{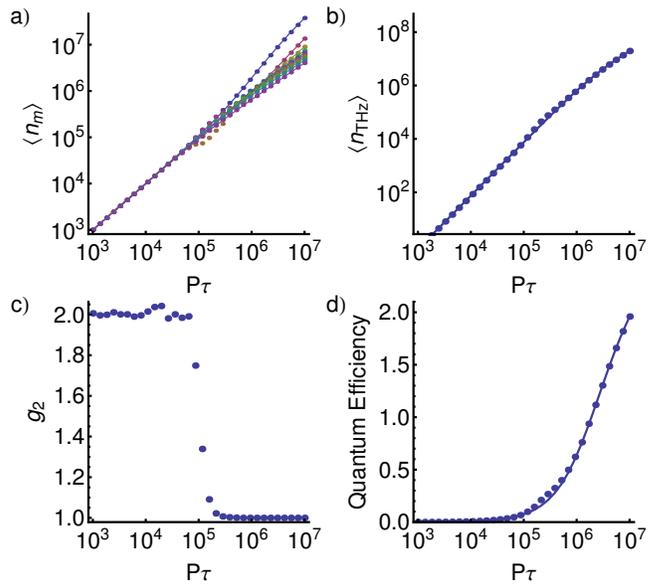}
\caption{(color online) Kinetic Monte Carlo simulation of a bosonic cascade laser with $M=10$ levels. a) Power dependence of the level occupations. b) Power dependence of the THz mode occupation. c) Power dependence of the second order correlation function of the THz mode. d) Power dependence of the quantum efficiency of THz generation. In all panels the points show results from kinetic Monte Carlo simulation, while the solid curves in (a), (b), and (d) show the result of solution of the classical Boltzmann equations (see Appendix).}
\label{fig:Quantum Cascade}
\end{figure}

Figure~\ref{fig:Quantum Cascade} shows results from the kinetic Monte Carlo modelling of a bosonic cascade of $M=10$ levels. The mode occupations agree fully with the result from the corresponding classical Boltzmann rate equations, however, the kinetic Monte Carlo approach provides an additional access to the second order correlation function. This reveals the smooth transition from an incoherent to a coherent state of THz photons in the THz cavity.

For completeness we also calculated the quantum efficiency of THz emission, which, we define as the ratio of the number of THz photons emitted by the system to the number of optical frequency photons put into the system~\cite{Liew2013}. This is equivalent to the ratio of the THz photon emission rate ($n_T/\tau_T$) to the total system pumping rate ($MP_0$):
\begin{equation}
Q=\frac{n_T}{M\tau_TP_0}
\end{equation}
The cascade geometry allows THz lasing to appear with quantum efficiency exceeding unity, in the regime of stimulated scattering despite the weak spontaneous scattering rates in the system.

\section{Conclusion}

The kinetic Monte Carlo approach offers an efficient numerical simulation of quantum Boltzmann equations, suitable for the description of non-equilibrium bosonic systems. Such an approach offers access to the second order correlation function and allows to study its development across non-equilibrium phase transitions. As an example, we considered the behaviour of the second order correlation function in one-dimensional exciton-polariton microwires. Here we are able to treat the cross-over from an incoherent state, to a non-ground state, to a ground quasi-condensate. While we required interactions between polaritons and hot excitons to provide a mechanism of energy relaxation from first principles, interactions between the condensing polaritons themselves were not required to generate the condensate. We also studied the formation of coherence in a bosonic cascade laser geometry, verifying the possibility of coherent terahertz emission with quantum efficiency exceeding unity.

We hope that the kinetic Monte Carlo approach can also serve in the description of partial energy relaxation and coherence formation in a variety of other exciton-polariton systems. In particular, we anticipate that the quantum optics of geometries confined with static potentials~\cite{Lai2007,Kim2013,Ostrovskaya2013,Zhang2015,Winkler2016} and self-induced traps~\cite{Askitopoulos2015,Gao2015b}, where transitions were observed between non-ground and ground quasi condensates would be accessible.

{\it Acknowledgements:} T.L. was supported by the
Singaporean Ministry of Education -- Singapore Academic Research Fund Tier-2 project (Project No. 2015-T2-1-055) and Tier-1 project (Project No. 2016-T1-1-084). D.P. was supported by the Singaporean Ministry of Education -- Singapore Academic Research Fund Tier-2 project (Project No. MOE2014-T2-2-119, with WBS No. R-144-000-350-112). I.G.S. was supported by IBS-R024-D1, the Australian Research Council's Discovery Projects funding scheme (project DE160100167), and the President of Russian Federation (project MK-5903.2016.2).

\appendix

\section{Polariton Relaxation Mediated by Acoustic Phonons}
The phonon-assisted sctattering rate between two polariton states involving excitons of wavevector ${\bf{k}}_1$ and ${\bf{k}}_2$ is computed following Ref.\cite{Piermarocchi1996} as
\begin{eqnarray}
\label{eq:Wk1k2}
{W_{{{\bf{k}}_1} \to {{\bf{k}}_2}}} &=& \frac{{{L_z}}}{{\rho uV}}\frac{{{{\left| {\Delta {\bf{k}}} \right|}^2} + q_z^2}}{{\left| {\hbar u{q_z}} \right|}}{\left| {X\left( {{{\bf{k}}_1}} \right)X\left( {{{\bf{k}}_2}} \right)} \right|^2}\\
\nonumber &\times& {\left[ {{a_e}I_e^{||}\left( {\left| {\Delta {\bf{k}}} \right|} \right)I_e^ \bot \left( {{q_z}} \right) - {a_h}I_h^{||}\left( {\left| {\Delta {\bf{k}}} \right|} \right)I_h^ \bot \left( {{q_z}} \right)} \right]^2}
\end{eqnarray}
where $\Delta {\bf{k}} = {{\bf{k}}_1} - {{\bf{k}}_2}$ and $q_z$ is the projection of the phonon momentum on the vertical axis. We use here typical $GaAs$ material paramters: $\rho=5318$ kg/m$^3$ is the material density, $L_z=10$ nm and $V=\pi R^2 L_z$ are the quantum well thickness and volume (for a microcavity radius $R$) respectively. $u=3350$ m/s is the speed of sound, $a_{\rm{e}}=-7$ eV and $a_{\rm{h}}=2.7$ eV are the lattice deformation potentials of the induced by phonons at the locations of electrons and holes. ${I_{e,h}^{||}}$ and ${I_{e,h}^{\bot}}$ are the overlap integrals of the phonon wave functions with the electron and hole wave functions, respectively, in the in-plane and growth directions. They are expressed as
\begin{eqnarray}
I_{e,h}^{||}\left( {\left| {\Delta {\bf{k}}} \right|} \right) = {\left[ {1 + {{\left( {\frac{{{m_{e,h}}}}{{{m_{\rm{e}}} + {m_{\rm{h}}}}}|\Delta {\bf{k}}|{a_{\rm{B}}}} \right)}^2}} \right]^{ - 3/2}}\\
I_{e,h}^ \bot \left( {{q_z}} \right) = \frac{{{\pi ^2}}}{{\frac{{{q_z}{L_z}}}{2}\left[ {{\pi ^2} - {{\left( {\frac{{{q_z}{L_z}}}{2}} \right)}^2}} \right]}}{\rm{sin}}\left( {\frac{{{q_z}{L_z}}}{2}} \right)
\end{eqnarray}
where $m_\textrm{e}=0.067m_0$ and $m_\textrm{h}0.18m_0$ are the effective masses of electrons and holes in terms of the free electron mass $m_0$ and $a_\textrm{B}=10$ nm is the exciton Bohr radius. Finally $X({\bf k})$ is the excitonic fraction defined as
\begin{equation}
X\left( {\bf{k}} \right) = \frac{2}{{\sqrt {4 + {{\left| {{E_{\rm p}}\left( {\bf{k}} \right)/{\Omega _R}} \right|}^2}} }}
\end{equation}
where ${E_{\rm p}}\left( {\bf{k}} \right)$ is the polariton dispersion relation and $\Omega_R=10$ meV is the Rabi splitting.

\section{Polariton Relaxation Mediated by Hot Excitons}

The matrix elements of scattering between polariton and hot-exciton states are obtained from the Fermi Golden rule following Ref.~\onlinecite{Porras2002}. The scattering rate of the process illustrated in Fig.~\ref{fig:ScatteringMechanisms}b, from a polariton state of wavevector $k_1$ and exciton state of wavevector $k_3$ to a polariton state of wavevector $k_2$ and an exciton state of wavevector $k_4$, is given by:
\begin{equation}
W_{k_1,k_2}=\frac{2\pi}{\hbar}\left(\frac{Lm_X}{\pi\hbar^2\sqrt{|k_3k_4|}}\right)\left(\frac{6E_Ba_B^2}{S}\right)^2n_{ex,k_3}
\end{equation}
Here the factor in the first parenthesis on the right-hand side is an average of the initial and final exciton density of states in one-dimension (assuming a parabolic exciton dispersion). The factor in the second parenthesis is the matrix element of exciton-exciton scattering~\cite{Tassone1999}, with $E_B$ the exciton binding energy, $a_B$ the exciton Bohr radius and $S$ a normalization area. $n_{ex,k_3}$ is the occupation of excitons in initial state $k_3$. For each combination of wavevectors $k_1$ and $k_2$, $k_3$ and $k_4$ are obtained from energy and momentum conservation:
\begin{align}
k_3&=\frac{2m_X}{\hbar^2}\frac{E_{k_1}-E_{k_2}}{2(k_1-k_2)}-\frac{k_1-k_2}{2}\\
k_4&=k_1-k_2+k_3
\end{align}
where $E_k$ represents the polariton dispersion.

Similar expressions can be used for the processes in Figs.~\ref{fig:ScatteringMechanisms}c and d. We note that in the case of Fig.~\ref{fig:ScatteringMechanisms}c one should sum over a few different processes that can satisfy the energy and momentum (phase matching) conditions.

\section{Classical Boltzmann Equations for the Bosonic Cascade}

In the classical regime, the quantum cascade can be modelled by a set of classical rate equations~\cite{Liew2013} for the mode occupations, $n_\lambda$
\begin{align}
\frac{dn_M}{dt}&=P_0+W\left(n_{M-1}(n_M+1)n_T\right.\notag\\
&\hspace{20mm}\left.-n_M(n_{M-1}+1)(n_T+1)\right)-\frac{n_M}{\tau}\\
\frac{dn_\lambda}{dt}&=P_0+W\left(n_{\lambda-1}(n_\lambda+1)n_T\right.\notag\\
&\hspace{20mm}\left.-n_\lambda(n_{\lambda-1}+1)(n_T+1)\right.\notag\\
&\hspace{20mm}\left.+n_{\lambda+1}(n_\lambda+1)(n_T+1)\right.\notag\\
&\hspace{20mm}\left.-n_\lambda(n_{\lambda+1}+1)n_T\right)-\frac{n_\lambda}{\tau}\\
\frac{dn_1}{dt}&=P_0+W\left(n_2(n_1+1)(n_T+1)\right.\notag\\
&\hspace{20mm}\left.-n_1(n_2+1)n_T\right)-\frac{n_1}{\tau}
\end{align}
where $1<\lambda<M$, and the THz mode occupation $n_T$,
\begin{align}
\frac{dn_T}{dt}&=W\sum_\lambda^2\left(n_\lambda(n_{\lambda-1}+1)(n_T+1)\right.\notag\\
&\hspace{20mm}\left.-n_{\lambda-1}(n_\lambda+1)n_T\right)-\frac{n_T}{\tau_T}
\end{align}
Equations for the steady state are readily obtained by setting the time derivatives to zero:
\begin{align}
n_M&=\frac{P_0\tau+W\tau n_{M-1}n_T}{W\tau\left(n_T+1+n_{M-1}\right)+1}\\
n_\lambda&=\frac{P_0\tau+W\tau\left((n_{\lambda-1}+n_{\lambda+1})n_T+n_{\lambda+1}\right)}{W\tau\left(2n_T+1+n_{\lambda-1}-n_{\lambda+1}\right)+1}\\
n_1&=\frac{P_0\tau+W\tau\left(n_T+1\right)n_2}{W\left(n_T-n_2\right)+1}\\
n_T&=\frac{W\tau\sum_{\lambda=2}^M\left(1+n_{\lambda-1}\right)n_\lambda}{W\tau\left(n_1-n_M\right)+\frac{\tau}{\tau_T}}
\end{align}
A simultaneous solution to this set of equations can be easily found by starting from an initially unoccupied state and evaluating the quantities $n_\lambda$ and $n_T$ iteratively until the equations become consistent. The result of this procedure gives rise to the solid curves in Figs.~\ref{fig:Quantum Cascade}a, b, and d, which are in agreement with the result of full kinetic Monte Carlo modelling.


\begin{thebibliography}{10}

\bibitem{Klaers2010}
J Klaers, J Schmitt, F Vewinger, \& M Weitz, Nature, {\bf 468}, 545 (2010).

\bibitem{Kasprzak2006} J Kasprzak, M Richard, S Kundermann, A Baas, P Jeambrun, J M J Keeling, F M Marchetti, M H Szyma\'nska, R Andr\'e, J L Staehli, V Savona, P B Littlewood, B Deveaud, \& Le Si Dang, Nature, {\bf 443}, 409 (2006).

\bibitem{Balili2007} R Balili, V Hartwell, D Snoke, L Pfeiffer, \& K West, Science, {\bf 316}, 1007 (2007).

\bibitem{Lai2007} C W Lai, N Y Kim, S Utsunomiya, G Roumpos, H Deng, M D Fraser, T Byrnes, P Recher, N Kumada, T Fujisawa, \& Y Yamamoto, Nature, {\bf 450}, 529 (2007).

\bibitem{Byrnes2014}
T Byrnes, N Y Kim, \& Y Yamamoto, Nature Phys., {\bf 10}, 803 (2014).

\bibitem{Love2008}
A P D Love, D N Krizhanovskii, D M Whittaker, R Bouchekioua, D Sanvitto, S Al Rizeiqi, R Bradley, M S Skolnick, P R Eastham, R Andre, \& L S Dang, Phys. Rev. Lett., {\bf 101}, 067404 (2008).

\bibitem{Kasprzak2008}
J Kasprzak, M Richard, A Baas, B Deveaud, R Andr\'e, J-Ph Poizat, \& L S Dang, Phys. Rev. Lett., {\bf 100}, 067402 (2008).

\bibitem{Horikiri2010}
T Horikiri, P Schwendimann, A Quattropani, S H\"ofling, A Forchel, \& Y Yamamoto, Phys. Rev. B, {\bf 81}, 033307 (2010).

\bibitem{Tempel2012}
J-S Tempel, F Veit, M A\ss mann, L E Kreilkamp, A Rahimi-Iman, A L\"offler, S H\"ofling, S Reitzenstein, L Worschech, A Forchel, \& M Bayer, Phys. Rev. B, {\bf 85}, 075318 (2012).

\bibitem{RahimiIman2012}
A Rahimi-Iman, A V Chernenko, J Fischer, S Brodbeck, M Amthor, C Schneider, A Forchel, S H\"ofling, S Reitzenstein, \& M Kamp, Phys. Rev. B, {\bf 86}, 155308 (2012).

\bibitem{Amthor2015}
M Amthor, H Flayac, I G Savenko, S Brodbeck, M Kamp, T Ala-Nissila, C Schneider, \& S H\"ofling, arXiv: 1511.00878 (2015).

\bibitem{Kim2016}
S Kim, B Zhang, Z Wang, J Fischer, S Brodbeck, M Kamp, C Schneider, S Hofling, \& H Deng, Phys. Rev. X, {\bf 6}, 011026 (2016).

\bibitem{Maragkou2010} M Maragkou, A. J. D. Grundy, E. Wertz, A. Lema\^itre, I. Sagnes, P. Senellart, J. Bloch, \& P. G. Lagoudakis, Phys. Rev. B, {\bf 81}, 081307(R) (2010).

\bibitem{Vorberg2013}
D Vorberg, W Wustmann, R Ketzmerick, \& A Eckardt, Phys. Rev. Lett., {\bf 111}, 240405 (2013).

\bibitem{Porras2002}
D Porras, C Ciuti, J J Baumberg, \& C Tejedor, Phys. Rev. B, {\bf 66}, 085304 (2002).

\bibitem{Doan2005}
T D Doan, H T Cao, D B T Thoai, \& H Haug, Phys. Rev. B, {\bf 72}, 085301 (2005).

\bibitem{Kasprzak2008b}
J Kasprzak, D D Solnyshkov, R Andr\'e, Le Si Dang, \& G Malpuech, Phys. Rev. Lett., {\bf 101}, 146404 (2008).

\bibitem{Cao2008}
H T Cao, T D Doan, D B T Thoai, \& H Haug, Phys. Rev. B, {\bf 77}, 075320 (2008).

\bibitem{Read2009}
D Read, T C H Liew, Yu G. Rubo, \& A V Kavokin, Phys. Rev. B, {\bf 80}, 195309 (2009).

\bibitem{Wouters2010}
M Wouters, T C H Liew, \& V Savona, Phys. Rev. B, {\bf 82}, 245315 (2010).

\bibitem{Wouters2012}
M Wouters, New J. Phys., {\bf 14}, 075020 (2012).

\bibitem{Sieberer2013}
L M Sieberer, S D Huber, E Altman, \& S Diehl, Phys. Rev. Lett., {\bf 110},	195301 (2013).

\bibitem{Savenko2013}
I G Savenko, T C H Liew, \& I A Shelykh, Phys. Rev. Lett., {\bf 110}, 127402 (2013).

\bibitem{Solnyshkov2014}
D D Solnyshkov, H Ter\c{c}as, K Dini, \& G Malpuech, Phys. Rev. A, {\bf 89}, 033626 (2014).

\bibitem{Racine2014}
D Racine \& P R Eastham, Phys. Rev. B, {\bf 90}, 085308 (2014).

\bibitem{Molmer1993}
K Molmer, Y Castin, \& J Dalibard, J. Opt. Soc. Am. B, {\bf 10}, 524 (1993).

\bibitem{Flayac2015}
H Flayac, I G Savenko, M M\"ott\"onen, T Ala-Nissila, Phys. Rev. B, {\bf 92}, 115117 (2015).

\bibitem{Liew2013}
T C H Liew, M M Glazov, K V Kavokin, I A Shelykh, M A Kaliteevski, \& A V Kavokin, Phys. Rev. Lett., {\bf 110}, 047402 (2013).

\bibitem{Tzimis2015}
A Tzimis, A V Trifonov, G Christmann, S I Tsintzos, Z Hatzopoulos, I V Ignatiev, A V Kavokin, \& P G Savvidis, Appl. Phys. Lett., {\bf 107},	101101	(2015).

\bibitem{Chaturvedi1977}
S Chaturvedi, C W Gardiner, I S Matheson, \& D F Walls, J. Stat. Phys., {\bf 17}, 469 (1977).

\bibitem{Liew2016}
T C H Liew, Y G Rubo, A S Sheremet, S De Liberato, I A Shelykh, F P Laussy, \& A V Kavokin, New J. Phys., {\bf 18}, 023041 (2016).

\bibitem{Young1966}
W M Young \& E W Elcock, P Phys. Soc., {\bf 89}, 735 (1966).

\bibitem{Bortz1975}
A B Bortz, M H Kalos, \& J L Lebowitz, J. Comput. Phys., {\bf 17}, 10 (1975).

\bibitem{Gillespie1976}
D T Gillespie, J. Comput. Phys., {\bf 22}, 403 (1976).

\bibitem{Fichthorn1991}
K A Fichthorn \& W H Weinberg, J. Chem. Phys., {\bf 95}, 1090 (1991).

\bibitem{Laussy2008}
F P Laussy, E del Valle, \& C Tejedor, Phys. Rev. Lett., {\bf 101}, 083601 (2008).

\bibitem{Savvidis2000}
P G Savvidis, J J Baumberg, R M Stevenson, M S Skolnick, D M Whittaker, \& J S Roberts, Phys. Rev. Lett., {\bf 84}, 1547 (2000).

\bibitem{Keeling2008}
J Keeling \& N G Berloff, Phys. Rev. Lett., {\bf 100}, 250401 (2008).

\bibitem{Richard2005} M Richard, J Kasprzak, R Andr\'e, R Romestain, L S Dang, G Malpuech, \& A Kavokin, Phys. Rev. B, {\bf 72}, 201301 (2005).

\bibitem{Wertz2010} E Wertz, L Ferrier, D Solnyshkov, R Johne, D Sanvitto, A Lema\^itre, I Sagnes, R Grousson, A V Kavokin, P Senellart, G Malpuech, \& J Bloch, Nature Phys., {\bf 6}, 860 (2010).

\bibitem{Anton2012}
C Anton, T C H Liew, G Tosi, M D Mart\'in, T Gao, Z Hatzopoulos, P S Eldridge, P G Savvidis, \& L Vi\~na, Appl. Phys. Lett., {\bf 101}, 261116 (2012).

\bibitem{Anton2013}
C Anton, T C H Liew, G Tosi, M D Mart\'in, T Gao, Z Hatzopoulos, P S Eldridge, P G Savvidis, \& L Vi\~na, Phys. Rev. B, {\bf 88}, 245307 (2013)

\bibitem{Anton2015}
C Ant\'on, S Morina, T Gao, P S Eldrdige, T C H Liew, M D Mart\'in, Z Hatzopoulos, P G Savvidis, I A Shelykh, \& L Vi\~na, Phys. Rev. B, {\bf 91}, 075305 (2015).

\bibitem{Gao2015}
T Gao, C Ant\'on, T C H Liew, M D Mart\'in, Z Hatzopoulos, L Vi\~na, P S Eldridge, \& P G Savvidis, Appl. Phys. Lett., {\bf 107}, 011106 (2015).

\bibitem{Piermarocchi1996}
C Piermarocchi, F Tassone, V Savona, A Quattropani, \& P Schwendimann, Phys. Rev. B, {\bf 53}, 15834 (1996).

\bibitem{Sun2017}
Y Sun, P Wen, Y Yoon, G Liu, M Steger, L N Pfeiffer, K West, D W Snoke, \& K A Nelson, Phys. Rev. Lett., {\bf 118}, 016602 (2017).

\bibitem{Mouchliadis2008}
L Mouchliadis \& A L Ivanov, Phys. Rev. B, {\bf 78}, 033306 (2008).

\bibitem{Belykh2013}
V V Belykh, N N Sibeldin, V D Kulakovskii, M M Glazov, M A Semina, C Schneider, S H\"ofling, M Kamp, \& A Forchel, Phys. Rev. Lett., {\bf 110}, 137402 (2013).

\bibitem{Sarchi2007}
D Sarchi \& V Savona, Phys. Rev. B, {\bf 75}, 115326 (2007).

\bibitem{Doan2008}
T D Doan, H T Cao, D B T Thoai, \& H Haug, Phys. Rev. B, {\bf 78}, 205306 (2008).

\bibitem{Kim2013}
N Y Kim, K Kusudo, A Loffler, S Hofling, A Forchel, \& Y Yamamoto, New J. Phys., {\bf 15}, 035032 (2013).

\bibitem{Ostrovskaya2013}
E A Ostrovskaya, J Abdullaev, M D Fraser, A S Desyatnikov, \& Y S Kivshar, Phys. Rev. Lett., {\bf	110}, 170407 (2013).

\bibitem{Zhang2015}
L Zhang, W Xie, J Wang, A Poddubny, J Lu, Y Wang, J Gu, W Liu, D Xu, X Shen, Y G Rubo, B L Altshuler, A V Kavokin, \& Z Chen, Proc. Natl. Acad. Sci. U. S. A., {\bf 112}, E1516 (2015).

\bibitem{Winkler2016}
K Winkler, O A Egorov, I G Savenko, X Ma, E Estecho,T Gao, S M\"uller, M Kamp, T C H Liew, E A Ostrovskaya, S \"ofling, \& C Schneider, Phys. Rev. B, {\bf 93}, 121303(R) (2016).

\bibitem{Askitopoulos2015}
Askitopoulos, T C H Liew, H Ohadi, Z Hatzopoulos, P G Savvidis, \& P G Lagoudakis, Phys. Rev. B, {\bf 92}, 035305 (2015).

\bibitem{Gao2015b}
T Gao, E Estrecho, K Y Bliokh, T C H Liew, M D Fraser, S Brodbeck, M Kamp, C Schneider, S H\"ofling, Y Yamamoto, F Nori, Y S Kivshar, A G Truscott, R G Dall, \& E A Ostrovskaya,
Nature, {\bf 526}, 554 (2015).

\bibitem{Tassone1999}
F Tassone \& Y Yamamoto, Phys. Rev. B, {\bf 59}, 10830 (1999).

\end{thebibliography}
\end{document}